\documentclass[symmetry,article,accept,pdftex,moreauthors]{Definitions/mdpi} 
\firstpage{1} 
\pubvolume{15}
\issuenum{2}
\articlenumber{499}
\pubyear{2023}
\copyrightyear{2023}
\externaleditor{Academic Editors: { Qun Wang, Zuotang Liang, Enke Wang and Ignatios Antoniadis}}

\datereceived{23 December 2022} 
\daterevised{12 January 2023}
\dateaccepted{27 January 2023} 
\datepublished{13 February 2023} 
\hreflink{https://doi.org/10.3390/\linebreak sym15020499} 

\newcommand{\SNN}{$\sqrt{s_{NN}}$}
\usepackage{ragged2e}
\usepackage{wrapfig}
\usepackage{amsmath}

\Title{Collective Excitation in High-Energy Nuclear Collisions---In Memory of Professor Lianshou Liu}

\TitleCitation{Collective Excitation in High-Energy Nuclear Collisions---In Memory of Professor Lianshou Liu}

\Author{Huan Zhong Huang $^{1}$, Feng Liu $^{2}$, Xiaofeng Luo $^{2}$, Shusu Shi $^{2,}$*\orcidA{}, Fuqiang Wang $^{3}$ and Nu Xu $^{4}$}

\AuthorNames{Huan Zhong Huang, Feng Liu, Xiaofeng Luo, Shusu Shi, Fuqiang Wang and Nu Xu}

\AuthorCitation{Huang, H.Z.; Liu, F.; Luo, X.; Shi, S.; Wang, F.; Xu, N.}

\address{%
$^{1}$ \quad Department of Physics and Astrophysics, University of California, Los Angeles, CA 90095, USA\\

$^{2}$ \quad Key Laboratory of Quark \& Lepton Physics (MOE) and Institute of Particle Physics, Central China Normal University, Wuhan 430079, China\\
$^{3}$ \quad Department of Physics, Purdue University, West Lafayette, IN 47907, USA\\
$^{4}$ \quad Nuclear Science Division, Lawrence Berkeley National Laboratory, Berkeley, CA 94720, USA}

\corres{Correspondence: shiss@mail.ccnu.edu.cn}

\abstract{We celebrate the legacies of our friend and mentor Professor Lianshou Liu who was one of the pioneers for the phenomenology of multi-particle interactions and initiated the physics of relativistic heavy-ion collisions in China. 
In this article, we discuss some of the recent exciting experimental observations on the collective phenomena including collectivity, chirality, criticality, strangeness production, and thermal equilibrium in high-energy nuclear collisions. Future directions, especially the physics at high baryon density, will be discussed with a focus on the first-order phase boundary and hyperon--nucleon interactions. 
}


\begin{document}


\section{Collectivity: Azimuthal Angular Anisotropy in High-Energy Nuclear Collisions}
Collective flow defined by the coefficiencies of the Fourier expansion of final particle distribution in momentum space is sensitive to the early stage of nuclear collisions. Specifically, the first three coefficiencies are called directed flow ($v_1$), elliptic flow ($v_2$), and triangular flow ($v_3$), respectively. 
Directed flow is sensitive to the Equation of State (EoS) of the medium;
elliptic flow is sensitive to the degree of freedom, partonic or hadronic level, and degree of equilibrium of the medium;
triangular flow is sensitive to the initial geometry fluctuations. 
A comprehensive set of measurements have
been achieved in RHIC-STAR experiment of nuclear collisions~\cite{STAR:2022ncy, STAR:2017kkh, STAR:2015gge, STAR:2014clz, STAR:2017okv, STAR:2012och, STAR:2013ayu, STAR:2013cow, STAR:2015rxv}. 
The Number of Constituent Quark (NCQ) scaling of $v_n$ observed at high energy collisions ($>$20 GeV)
indicates that the partonic collectivity has been built-up~\cite{STAR:2022ncy, STAR:2017kkh, STAR:2015gge, STAR:2013cow, ALICE:2022wpn}. In particular, the $D$ meson also follows the NCQ scaling~\cite{STAR:2017kkh, ALICE:2022wpn, CMS:2017vhp}, suggesting that the 
charm quark collectivity is at the same level as that of $u$, $d$, and $s$ quarks; therefore, the created 
medium reaches (nearly) equilibrium.

\begin{figure}[H]
    \includegraphics[width=5.47in,keepaspectratio]{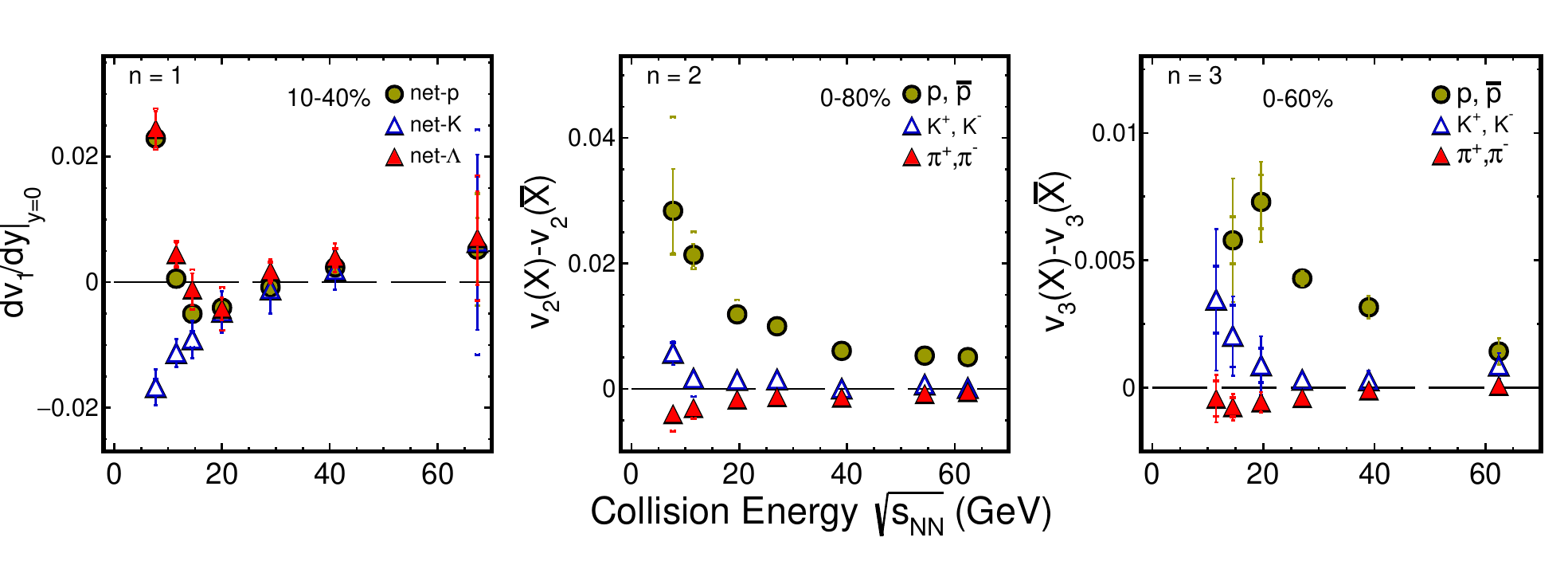}\vspace{-3pt}
    \caption{Collision energy dependence of directed flow slope at mid-rapidity (\textbf{left}),
    elliptic flow (\textbf{middle}), and triangular flow (\textbf{right}) difference of particles versus anti-particles. 
    The data are \mbox{from~\cite{STAR:2014clz, STAR:2017okv, STAR:2012och, STAR:2013ayu, STAR:2013cow, STAR:2015rxv, Parfenov:2020fuo}}.}
    \label{Fig:vn_energy} 

\end{figure}

The main motivation of Beam Energy Scan (BES) program is to explore the QCD phase diagram and search for the possible phase boundary and critical point. The first phase of BES program (BES-I) at STAR experiment covers collision energy $\sqrt{s_{NN}} = $ 7.7--62.4~GeV.
Lots of interesting phenomena have been observed; here, we focus on the collective flow $v_n$ measurements.
Figure~\ref{Fig:vn_energy} summarizes the directed, elliptic, and triangular flow relevant observations from STAR BES-I. The $v_1$ slope of net-baryon near mid-rapidity as a function of collision energy is regarded as a possible signal of first-order phase transition. The non-monotonic energy dependence of $v_1$ slope is associated with the phase transition and the minimum of $v_1$ slope is called the ``softest point collapse''~\cite{Stoecker:2004qu}. In the experiment, as neutrons are hard to be measured, we use net-proton as a proxy of net-baryon. The left panel of Figure~\ref{Fig:vn_energy} shows the $v_1$ slope of net-proton, net-Kaon, and net-$\Lambda$ as a function of collision energy. The non-monotonic behavior is observed for net-proton and net-$\Lambda$, and the minimum occurs near $\sqrt{s_{NN}}$ of 10--20 GeV. On the other hand, the net-Kaon slope shows a monotonic increase from low to high energy. Above $\sqrt{s_{NN}} = $ 20 GeV, the net-Kaon slope is overlapping with net-proton and net-$\Lambda$, while the divergence happens below 20 GeV. 
Further investigation is needed to understand the physical mechanism of this divergence at the low energy region. In addition, the centrality dependence measurements of net-particle $v_1$ slope are crucial to understand the effects from mechanism which is not related to phase transition~\cite{Nara:2022kbb, Nara:2021fuu, Nayak:2019vtn}. The $v_2$ of particles and corresponding anti-particles is also found to be significantly different below 20 GeV as shown in the middle panel of Figure~\ref{Fig:vn_energy}. Even though the observed NCQ scaling is not valid for particles and anti-particles, roughly scaling still works for particle group and anti-particle group separately. 
One explanation of the $v_2$ difference is the transported effect.
The final particles are a mixture of transported and produced particles, while the dominant part produces particles at high energy collisions, such as 200 GeV at RHIC and a few TeV at LHC. With the decrease of collision energy, there are more final particles from transportation. Since the transported particles undergo the whole collision evolution, the collectivity should be different with the produced particles. 
Several model studies try to explain the $v_2$ difference of particles and anti-particles and succeed partly~\cite{Dunlop:2011cf, Steinheimer:2012bn, Hatta:2015era, Xu:2013sta, Tu:2018ora, Liu:2019ags, Li:2020scf}. More detailed measurements as a function of centrality and for (multi-)strange hadrons help us pick out the right mechanism. A similar difference is also observed for $v_3$, as shown in the right panel of Figure~\ref{Fig:vn_energy}. The collision energy and particle type dependence are the same as $v_2$.
Recently, the $v_1$ and $v_2$ results of identified particles from $\sqrt{s_{NN}} = $ 3 GeV suggest that the hadronic interactions are dominant at such \mbox{energy~\cite{STAR:2021yiu, STAR:2021ozh, Lan:2022rrc}}.  
Comprehensive measurements of collective flow from collision energy $<$20~GeV, 
such as STAR experiment BES-II (3--20 GeV) and experiments at SPS and SIS18~\cite{Kuich:2021qog, NA49:2003njx, HADES:2020lob},
help us further constrain the phase transition boundary.

\vspace{-12pt}

\begin{figure}[H]
\includegraphics[width=5in,keepaspectratio]{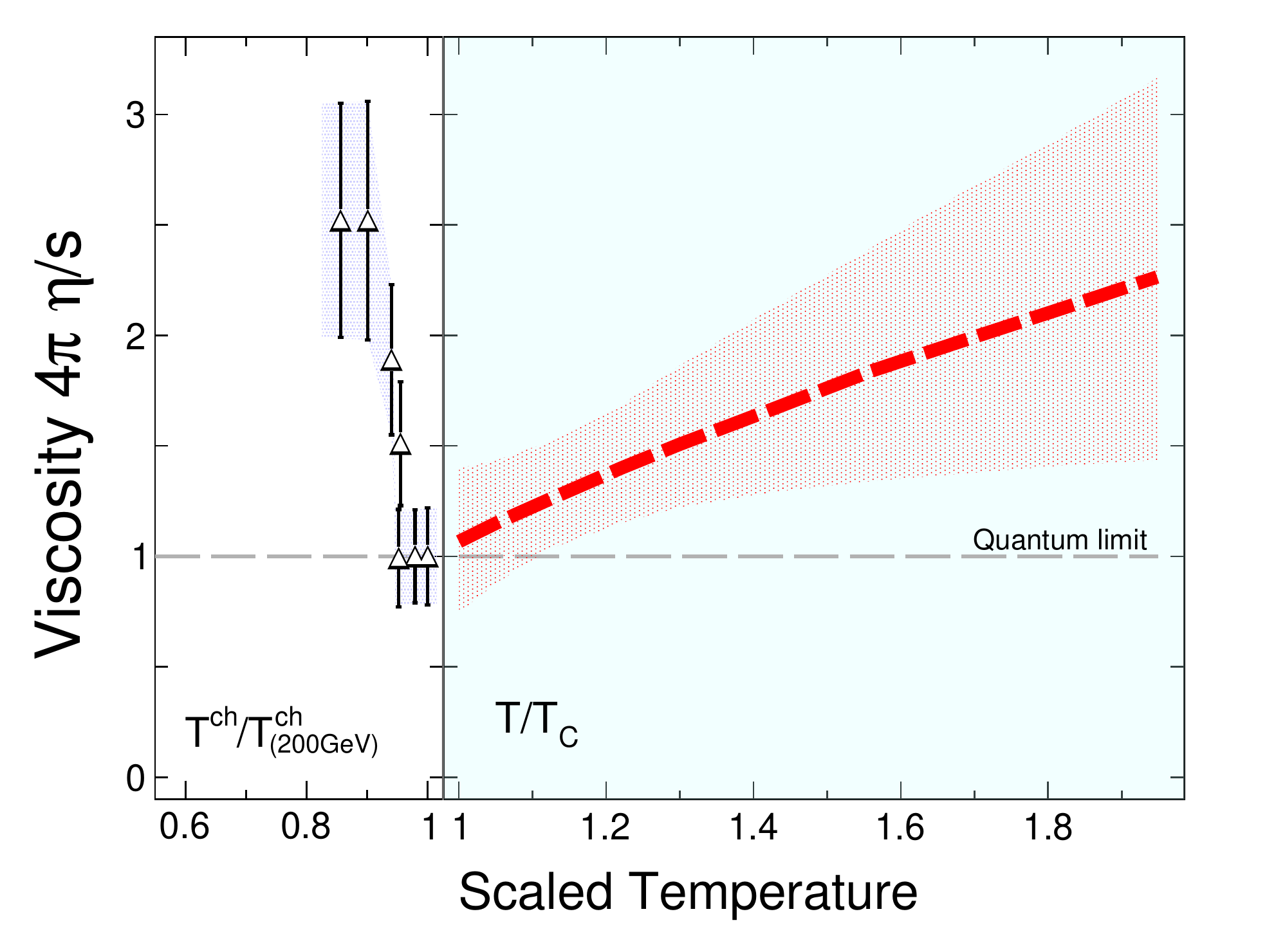}
\caption{The effective values of shear viscosity-to-entropy ratio, $4\pi \eta/s$, shown as a function of the scaled temperature. The horizontal dashed line indicates the quantum lower limit. Left panel: the extracted $4\pi \eta/s$ from the energy dependence of the measured $v_2$~\cite{STAR:2012och} and $v_3$~\cite{STAR:2016vqt}, shown as the scaled chemical freeze-out temperature ${\rm T_{ch}/T_{ch}}$ (200 GeV). Right panel: temperature evolution of $4\pi \eta/s$, extracted from Bayesian analyses~\cite{Bernhard:2019bmu,Xu:2017obm}.}\label{fig:etaos}
\end{figure}

The energy dependent experimental data on the collective flow can be used to extract important properties of the medium.
The temperature dependence of the event-averaged shear viscosity-to-entropy ratio $4\pi\eta/s$~\cite{Karpenko:2015xea} is shown in Figure~\ref{fig:etaos}. In the left panel, chemical freeze-out temperature from each energy~\cite{STAR:2017sal} is used and normalized to that from 200 GeV Au+Au collisions. As one can see, in the high energy limit, $\sqrt{s_{NN}}=$ 39--200 GeV, the ratio reaches unity, the quantum limit, implying that the medium created in such collisions is dominated by partonic interactions with a minimum value of $4\pi \eta/s$. At lower collision energies, on the other hand, hadronic interactions are dominant, and the medium shows a rapid increase of the viscosity-to-entropy ratio. The right panel is taken from Ref.~\cite{Karpenko:2015xea}, where the temperature evolution of the shear ratio is shown as a function of the scaled temperature $T/T_{c}$. Here, $T_{c}$ represents the critical temperature in the calculation~\cite{Bernhard:2019bmu,Xu:2017obm}.  The observed $V$-shaped feature is quite similar to what is described in Ref.~\cite{Csernai:2006zz} and can be taken as experimental evidence of the expected crossover transition in QCD.      

\section{Chirality: Chiral Magnetic Effect in High-Energy Nuclear Collisions}

The QCD vacuum can exist in states of varying topological charge or Chern--Simons winding number~\cite{Lee:1974ma,Morley:1983wr,Kharzeev:1998kz}. Sphaleron transitions between those states, strongly suppressed under normal conditions, may have appreciable probability at high temperatures, such as those achieved in high-energy nuclear collisions. Local metastable domains of gluon fields with non-zero topological charges may thus form in those collisions~\cite{Kharzeev:2004ey}. Interactions of quarks with those gluon fields can cause an imbalance in chirality, which may manifest as an electric current along a strong magnetic field, a phenomenon called the Chiral Magnetic Effect (CME)~\cite{Fukushima:2008xe}. 
Because those local domains break the parity ($\mathcal{P}$) and charge--parity ($\mathcal{CP}$) symmetries and because those domains arise ultimately from vacuum fluctuations, out of which our universe is believed to come into being, an observation of the CME in high-energy nuclear collisions may unravel the mystery of the matter--antimatter asymmetry of our universe~\cite{Dine:2003ax}. 

\textls[-25]{The CME has been extensively sought at the RHIC at BNL and the LHC at CERN~\cite{Kharzeev:2015znc,Zhao:2019hta}.} A signature of the CME is back-to-back charge separation along the magnetic field, produced mainly by spectator protons in relativistic heavy-ion collisions~\cite{Skokov:2009qp}. Because the magnetic field is on average perpendicular to the reaction plane (RP), a commonly used observable is the three-point azimuthal correlator~\cite{Voloshin:2004vk},
\vspace{-6pt}
\begin{equation}
    \Delta\gamma=\gamma_{\rm OS}-\gamma_{\rm SS}\;\;\;\mbox{and}\;\;\;
    \gamma=\langle\cos(\phi_\alpha+\phi_\beta-2\psi_{\rm RP})\rangle\,
\end{equation}
where $\phi_\alpha$ and $\phi_\beta$ are the azimuthal angles of particles $\alpha$ and $\beta$ of either the opposite sin (OS) or the same sign (SS).
The $\psi_{\rm RP}$ is the azimuthal angle of the RP and is usually reconstructed from final-state particles, whose inaccuracy is corrected by a resolution factor~\cite{Poskanzer:1998yz}. Equivalently, $\gamma$ can also be calculated by the three-particle correlator\linebreak $\gamma=\langle\cos(\phi_\alpha+\phi_\beta-2\phi_c)\rangle/v_{2,c}$, where $v_{2,c}$ is the elliptic flow parameter of the third particle $c$. Several other observables have been proposed and were found to be similar to $\Delta\gamma$~\cite{Choudhury:2021jwd}.

Charge separation measurements revealed strong $\Delta\gamma$ signals~\cite{Abelev:2009ac,Abelev:2009ad,Abelev:2012pa,Adamczyk:2013hsi}, approximately independent of collision energy except at lower RHIC energies where the signal dies off~\cite{Adamczyk:2014mzf}. However, major backgrounds exist that arise from two-particle correlations coupled with the finite elliptic flow of those background correlation sources~\cite{Wang:2009kd,Bzdak:2009fc,Schlichting:2010qia},
\vspace{-6pt}
\begin{equation}
    \Delta\gamma_{\rm bkgd}=\frac{N_{\rm 2p}}{N_{\alpha}N_{\beta}}\langle\cos(\phi_{\alpha}+\phi_{\beta}-2\phi_{\rm 2p})\rangle v_{2,{\rm 2p}}\,.
    \label{eq:bkgd}
\end{equation}
Here, $N_{\rm 2p}$ and $v_{\rm 2,2p}$ refer, respectively, to the number and elliptic flow of those two-particle correlation sources, such as resonances and jets, and $\phi_{\rm 2p}$ refers to their azimuths; $N_{\alpha,\beta}$ are the $\alpha,\beta$ particle multiplicities.
Experimental indication of major backgrounds comes from small-system measurements~\cite{Khachatryan:2016got,STAR:2019xzd}. Clear experimental evidence of those backgrounds is observed by the STAR measurement of $\Delta\gamma$ as a function of pair invariant mass~\cite{Zhao:2017nfq,STAR:2020gky}. 

It became clear because backgrounds are dominant that data-driven methods must be invoked to reliably extract the possible CME signal. An initial attempt was completed by STAR~\cite{Adamczyk:2013kcb} by analyzing data as a function of the event-by-event ellipticity of the particles of interest (i.e., $\alpha$ and $\beta$). 
A similar event-shape engineering (ESE) analysis~\cite{Schukraft:2012ah} was carried out by ALICE~\cite{Acharya:2017fau} and CMS~\cite{Sirunyan:2017quh} by analyzing data as a function of $v_2$ of the particles of interest at midrapidity in events selected according to the ellipticity in the forward-rapidity region. Data are consistent with vanishing CME signals and upper limits, on the order of 20\% of the inclusive $\Delta\gamma$ measurements, have been extracted. 

\begin{figure}[H]
\includegraphics[width=0.99\textwidth]{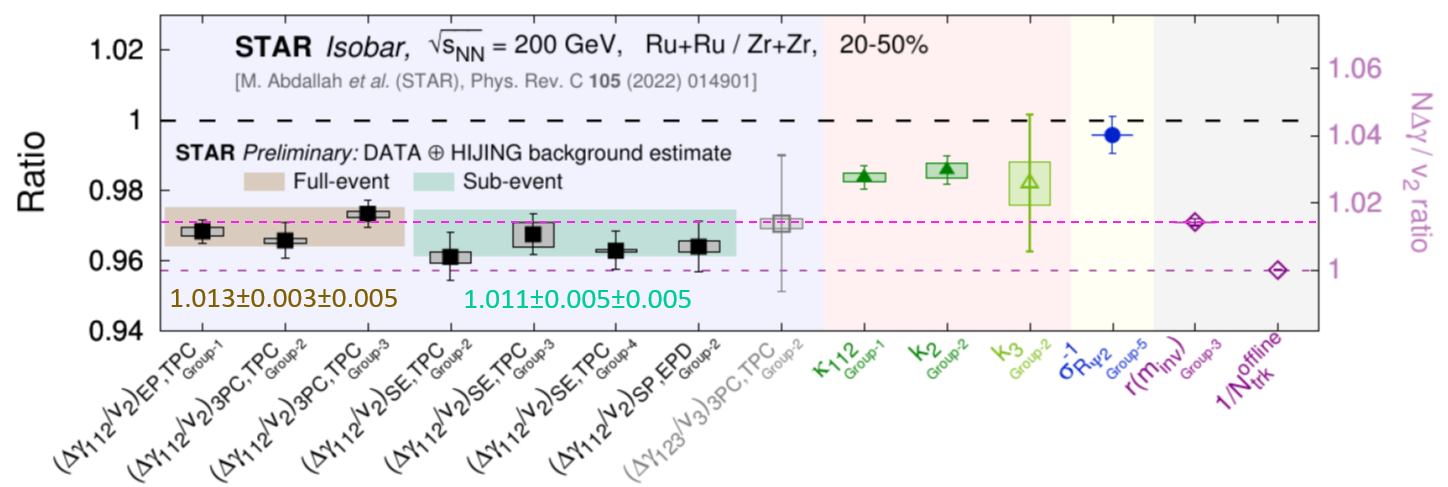}
\caption{The seven leftmost black data points show the Ru+Ru/Zr+Zr ratios of $\Delta\gamma/v_2$ (left coordinate) and $N\Delta\gamma/v_2$ (right coordinate) from STAR~\cite{STAR:2021mii}. The bottom and middle dashed lines indicate background baselines if the background scales with inverse multiplicity or relative pair excess $r=(N_{\rm OS}-N_{\rm SS})/N_{\rm OS}$, respectively. The shade bands indicate preliminary estimates of background baselines taking into account effects of nonflow contamination~\cite{Tribedy:QM,Wang:2022eoo,Feng:2022yus}.} 
\label{fig:isobar}
\end{figure}

To eliminate large backgrounds, experiments often rely on comparative or relative measurements. One such relative measurement is the isobar $^{44}_{96}$Ru+$^{44}_{96}$Ru and $^{40}_{96}$Zr+$^{40}_{96}$Zr collisions conducted in 2018 at RHIC~\cite{Skokov:2016yrj}. Because of the equal mass numbers of the isobar nuclei, the physics backgrounds are expected to be the same; the 10\% difference in the atomic numbers is expected to result in a difference in the magnetic fields and thus in the CME signal strengths between the two systems~\cite{Voloshin:2010ut}. A blind analysis was conducted~\cite{STAR:2019bjg}, and the results~\cite{STAR:2021mii} are shown in Figure~\ref{fig:isobar} with an unprecedented precision of 0.4\%. The Ru+Ru/Zr+Zr ratios of $\Delta\gamma/v_2$ (motivated by Equation~\eqref{eq:bkgd}) are all below unity, which has been understood to be due mainly to the multiplicity difference between the two isobar systems. This difference was predicted by energy density functional theory calculations~\cite{Xu:2017zcn,Li:2018oec} to arise from the smaller size of the $^{44}_{96}$Ru nucleus compared to that of the $^{40}_{96}$Zr nucleus~\cite{Xu:2021vpn}. If the $\Delta\gamma$ is inversely proportional to multiplicity, then the baseline for the double ratio would be the bottom dashed line in Figure~\ref{fig:isobar}. The measured double ratios are all above this line, seemingly suggesting finite CME signals, as also pointed out in Ref.~\cite{Kharzeev:2022hqz}. However, the inverse multiplicity scaling is only approximate because the background in $\Delta\gamma$ scales with the number of correlation sources (see Equation~\eqref{eq:bkgd}), which may not be strictly proportional to multiplicity. A more realistic baseline may be the Ru+Ru/Zr+Zr ratio of the pair excess $r=(N_{\rm OS}-N_{\rm SS})/N_{\rm OS}$~\cite{STAR:2021mii} indicated by the middle dashed line. If so, then there is no evidence of a CME signal in the isobar data. To complicate the matter further,  background contamination beyond that of Equation~\eqref{eq:bkgd} exists because of genuine three-particle correlations and nonflow contamination in $v_2$~\cite{Feng:2021pgf}. Preliminary estimates~\cite{Feng:2022yus} of those nonflow effects are shown by the shaded bands in Figure~\ref{fig:isobar}, indicating the final baselines for the double ratios measured by the full event and subevent methods. The isobar data are consistent with these baselines within approximately one standard deviation, suggesting that the CME signal in isobar collisions cannot be larger than a few percent---the 0.4\% data precision translates into a $f_{\rm CME}=2$--3\% CME fraction of the inclusive $\Delta\gamma$ measurement.

\begin{figure}[H]
\includegraphics[width=0.95\textwidth]{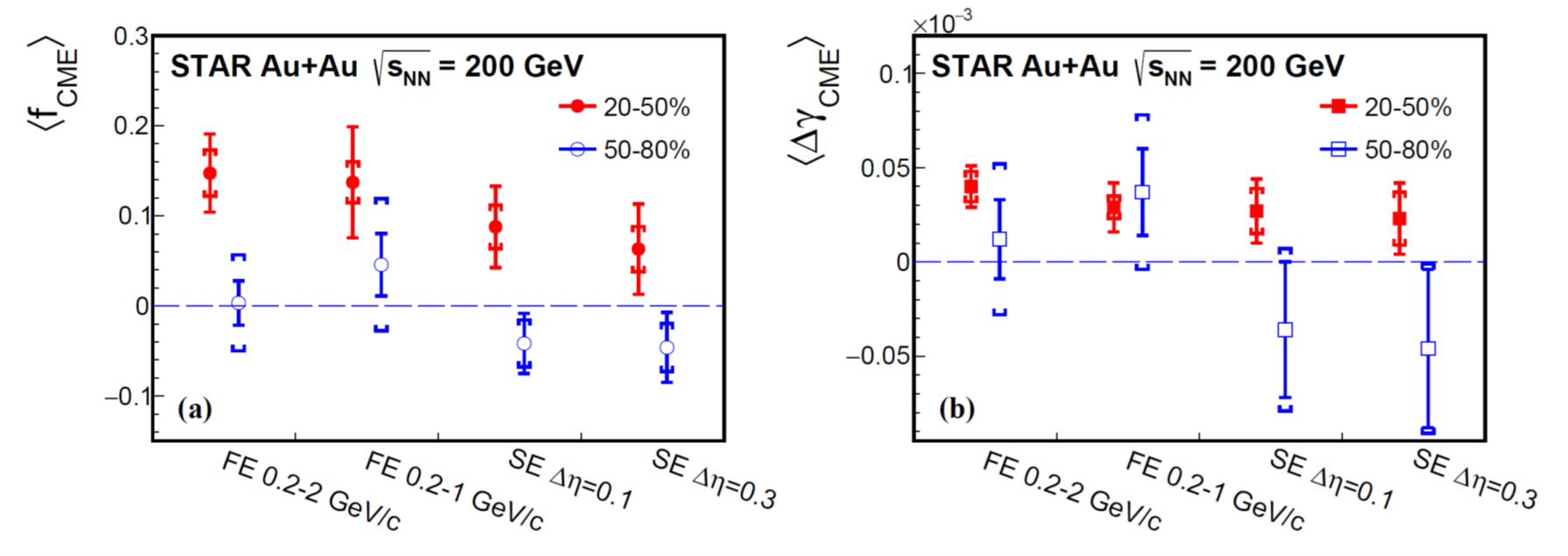}
\caption{The extracted CME signal fraction $f_{\rm CME}$ (\textbf{a}) and magnitude $\Delta\gamma_{\rm CME}$ (\textbf{b}) in the inclusive $\Delta\gamma$ measurements by the full-event (FE) and subevent (SE) methods in Au+Au collisions at 200~GeV from STAR~\cite{STAR:2021pwb}.}
\label{fig:pprp}
\end{figure}

Another comparative method consists of measurements with respect to the spectator plane (SP) and the participant plane (PP)~\cite{Xu:2017qfs,Voloshin:2018qsm}. Because the SP is better aligned with the perpendicular orientation of the magnetic field and the PP is determined by the elliptic flow harmonic direction, $\Delta\gamma$ measurements w.r.t.~SP and PP contain different contributions from the CME signal and flow background~\cite{Xu:2017qfs}. One can obtain the CME fraction by
$f_{\rm CME}\equiv\frac{\Delta\gamma_{\rm CME}\{\rm PP\}}{\Delta\gamma\{\rm PP\}}=\frac{A/a-1}{1/a^2-1}$, where $a$ is the ratio of $v_2$ measurement w.r.t.~SP to that w.r.t.~PP, which quantifies the angular spread between SP and PP, and $A$ is the ratio of $\Delta\gamma$ measurement w.r.t.~SP to that w.r.t.~PP. 
Figure~\ref{fig:pprp} shows the $f_{\rm CME}$ and the absolute $\Delta\gamma_{\rm CME}$ signal strength in Au+Au collisions at 200~GeV by STAR~\cite{STAR:2021pwb}. While the peripheral 50--80\% data are consistent with vanishing CME, the midcentral 20--50\% results indicate a finite CME signal of the order of 10\%, with a significance ranging from 1 to 3 standard deviations. While the flow-induced background is removed, the results are still contaminated by nonflow contributions. Model estimates~\cite{Feng:2021pgf} of those nonflow contributions indicate a consistent-with-zero contribution to the full-event result and a negative contribution to the subevent results. Estimation utilizing real data is ongoing. If nonflow contamination is small or even negative, then the positive $f_{\rm CME}$ observation by STAR may indeed indicate a finite CME signal. Note this does not contradict the null signal from the isobar data because the signal to background ratio in isobar collisions is expected to be significantly smaller, perhaps by a factor of three~\cite{Feng:2021oub}, than in Au+Au collisions, due to the smaller multiplicity (thus larger background) and the smaller magnetic field strength (thus smaller CME strength) in isobar collisions~\cite{Feng:2021oub}; the $f_{\rm CME}\sim10\%$ signal in Au+Au collisions and the less than a few percent signal in isobar collisions are, in fact, consistent with each other. 
\vspace{-6pt}

In summary, the CME-sensitive $\Delta\gamma$ measurements in heavy-ion collisions are dominated by flow-induced backgrounds arising from two-particle correlations coupled with elliptic flow $v_2$. Further background contamination arises from three-particle correlations and nonflow effects in $v_2$. Methods have been developed to eliminate those backgrounds, including event-shape engineering, isobar collisions, and measurements w.r.t.~spectator plane SP and participant plane PP. While the former two have yielded null CME signal with the present statistics, the comparative measurements w.r.t.~SP and PP indicate a hint of $f_{\rm CME}\sim10\%$ CME signal in Au+Au collisions at 200 GeV, with a significance of 1--3~standard deviations. 
In the future, an 
order of magnitude increase in Au+Au statistics is anticipated from 2023 and 2025 by STAR. This would present a powerful data set to either identify the CME or put a stringent upper limit on it. Such an outcome would significantly advance our understanding of QCD and our universe. 

\section{Criticality: Search for the QCD Critical Point and the Limit of Thermalization in Heavy-Ion Collisions}

\subsection{High Order Moments and Search for the QCD Critical Point}\label{sec:cp}
Exploring the phase structure of hot and dense QCD matter is one of the main goals of heavy-ion collisions. In the large baryon density region, the QCD critical point is the endpoint of the first-order phase transition boundary and a landmark in the QCD phase diagram. Mapping out the first-order phase boundary and pinning down the location of the QCD critical point will enhance our understanding of universe evolution and the structure of visible matters. During the last two decades, many efforts have been made both \linebreak experimentally~\cite{NA49:2012ebu,Czopowicz:2022nzy,STAR:2010mib,STAR:2013gus,STAR:2014egu,STAR:2017tfy,STAR:2020tga,STAR:2021iop,STAR:2021rls,STAR:2021fge,STAR:2022hbp,STAR:2022vlo} and theoretically~\cite{Stephanov:2006zvm,Fukushima:2010bq,Stephanov:2011pb,Stephanov:2017ghc,Bazavov:2017dus,Fu:2019hdw,Fu:2021oaw,Bzdak:2019pkr,Bazavov:2020bjn,Luo:2022mtp} to determine the location of the QCD critical point. In this review, we will focus on the measurement of high-order moments with the data taken in the first phase of the beam energy scan program at RHIC. 

For a quantum system, the thermodynamic pressure~\cite{Gavai:2004sd} can be expressed via temperature $T$ and chemical potentials $\mu_i$, where ${\rm (}i=B, Q, S{\rm )}$ stands for conserved quantum numbers of baryon ($B$), electric charge ($Q$), and strangeness ($S$) in heavy-ion collisions:
\begin{equation}
    P(T,\mu_B,\mu_Q,\mu_S,V) = \frac{T}{V}\sum_i{\rm ln}Z_i = \sum_i\pm\frac{Tg_i}{2\pi^2}\int{k^2dk{\rm ln}\{1\pm {\rm exp}[(\mu_i-E)/T]}\}.
\label{eq:pressure}
\end{equation}
Here, $k$, $g_i$, and $\pm$ signs stand for the particle momentum, the degeneracy, and the nature of Fermion ($+$) or Boson ($-$). The sum goes over all particles in the equilibrium with masses up to 2.5 GeV from the PDG. 

For a conserved quantity, the cumulants in thermal equilibrium can be expressed through the generalized thermodynamic susceptibilities:
\begin{equation}
    C_n^i=\frac{V}{T}T^n\chi_i^{(n)}
\end{equation}
and the $n^{\rm th}$ order generalized susceptibilities are the derivatives of the pressure:
\begin{equation}
    \chi_i^{(n)}=\frac{d^nP}{d\mu_i^n}
\end{equation}

Fluctuations and correlations among conserved charges, i.e., baryon number ($B$), electric charge ($Q$)
and strangeness ($S$), are sensitive observables to probe the QCD phase structure. Experimental proxies, such as net-protons, net-kaons, are used for measurements of mean ($M$), variance ($\sigma^{2}$), skewness ($s$), and kurtosis ($\kappa$)
of conversed charges~\cite{STAR:2010mib,STAR:2013gus,STAR:2017tfy}. As an example, the connections between moments and cumulant ($C_{n}$) ratios of conserved charges are listed as the following:\vspace{-6pt}
\begin{align}
\frac{\sigma^2_i}{M_i}&= \frac{C_2^i}{C_1^i} = \frac{\chi_2^i(T,\boldsymbol{\mu_i})}{\chi_1^i(T,\boldsymbol{\mu_i})}\label{eq:ratio0},\\
s_i\,\sigma_i&= \frac{C_3^i}{C_2^i} = \frac{\chi_3^i(T,\boldsymbol{\mu_i})}{\chi_2^i(T,\boldsymbol{\mu_i})}\label{eq:ratio1},\\
\kappa_i\, \sigma_i^2&=\frac{C_4^i}{C_2^i} = \frac{\chi_4^i(T,\boldsymbol{\mu_i})}{\chi_2^i(T,\boldsymbol{\mu_i})}\label{eq:ratio2}
\end{align}
where $i=B, Q$ and $S$ are the conserved charges, $C_n^i$ is the $n^{\rm th}$ order cumulant, and $\chi_n^i(T,\mu_i)$ is the generalized susceptibility defined as the $n^{\rm th}$ order derivative of pressure with respect to chemical potential $\boldsymbol{\mu_i}$.

One of the most important advantages of Equations~\eqref{eq:ratio0}--\eqref{eq:ratio2} is that they connect the measurements, the experimentally measured moments on the left sides of the equations, to the ratios of the thermodynamic susceptibilities (and ratios of cumulants) from theoretical calculations~\cite{Cheng:2008zh,Gavai:2010zn,Bazavov:2012vg,Borsanyi:2014ewa,Bellwied:2013cta,Fu:2021oaw}. These constructions offer not only a sensitive probe for studying the QCD phase structure in high-energy collisions but also to test the limits of thermalization in such collisions. The latter is a necessary step for us to understand the emerging macroscopic features from the violent microscopic scatterings.

\begin{figure}[H]
\includegraphics[width=0.9\textwidth]{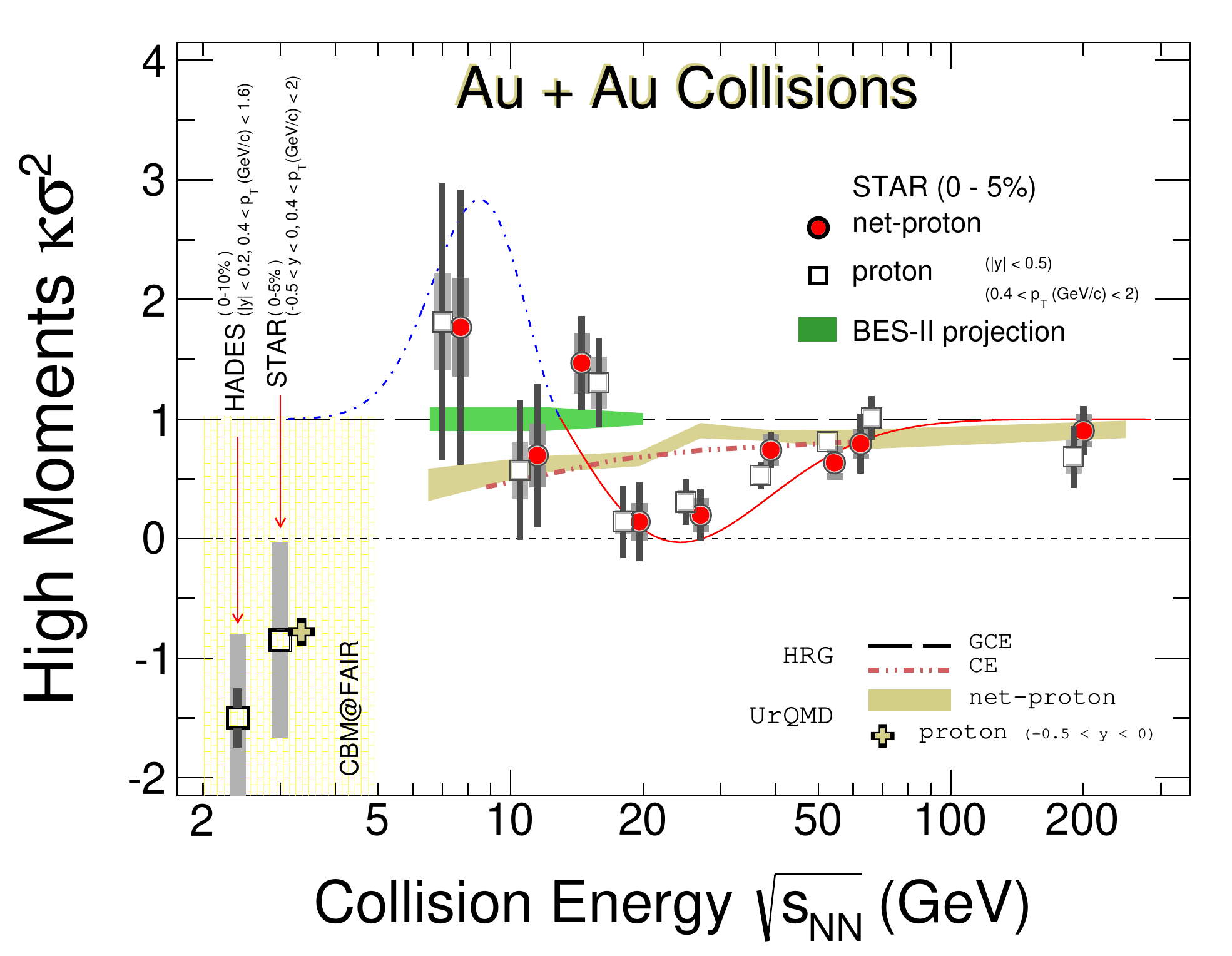}
\caption{Collision energy dependence of the $\kappa \sigma^{2}$ ($C_4/C_2$), for proton (squares) and net-proton (red circles) from top 0--5\% Au+Au collisions at RHIC~\cite{STAR:2021iop,STAR:2021fge}. The points for protons are shifted horizontally for clarity. The new result for proton from $\sqrt{s_{NN}}$ = 3\,GeV collisions is shown as a filled square. HADES data~\cite{HADES:2020wpc} of $\sqrt{s_{NN}}$ = 2.4\,GeV 0--10\% collisions are also shown. Statistical uncertainties are presented by bars while the systematic ones are indicated by the vertical grey bands. The green band is the estimated statistical uncertainties from the BES-II program. Results from the Hadron Resonance Gas model~\cite{Braun-Munzinger:2020jbk} and transport model UrQMD~\cite{Bleicher:1999xi,Bass:1998ca} are shown as the red dot dashed line and gold band, respectively. The thin solid red and dot-dashed blue lines depict a qualitative prediction for the behavior of the fourth-order net-proton cumulant due to an evolution in the vicinity of a critical region~\cite{Stephanov:2011pb}. }\label{BESI_netp}
\end{figure}

We start first from experimental observations. As a summary, Figure~\ref{BESI_netp} shows recent results on the fourth-order net-proton (filled red circles) and proton (open squares) high moments in central Au+Au collisions from the RHIC BES program and the HADES experiment~\cite{STAR:2020tga,STAR:2021fge,HADES:2020wpc} together with model comparisons. The thin solid red and dot-dashed blue lines depict a qualitative prediction for the behavior of the fourth-order net-proton cumulant ratio $C_{4}/C_{2}$ ($\kappa \sigma^{2}$) due to an evolution in the vicinity of a critical region~\cite{Stephanov:2011pb}; the locations of the peak of the dot-dashed blue curve and the dip of the solid red curve 
are chosen only for a qualitative comparison. 
Intriguing non-monotonic behavior is observed in the data, and while large error bars prevent one from making a more decisive conclusion, the behavior of the fourth-order net-proton cumulant ratio in the energy range \mbox{$\sqrt{s_{NN}}$ = 7.7--27\,GeV} seems to significantly deviate from the non-critical baseline provided by models. In particular, both the hadronic transport model UrQMD (cascade mode)~\cite{Bass:1998ca,Bleicher:1999xi} (gold band) and a thermal model in the canonical ensemble~\cite{Braun-Munzinger:2020jbk} (dot-dashed red line) predict, for decreasing collision energy, a monotonic suppression in the fourth-order cumulant ratio $C_{4}/C_{2}$ due to baryon number conservation, in contrast to the behavior tentatively seen in data. In addition, the experimental data at $\sqrt{s_{NN}} \le 20$\,GeV indicate an excess of two-proton correlations as compared to a non-critical baseline including effects due to, e.g., baryon number conservation~\cite{Vovchenko:2021kxx}. At 3 GeV, the agreement between data and transport model calculation, see the cross in the figure, implies that hadronic interaction dominates the properties of the medium~\cite{STAR:2021fge,STAR:2022qmt}. These data imply that the QCD critical region, if created in heavy-ion collisions, could only exist at energies higher than 3 GeV. 

The green band, in Figure~\ref{BESI_netp}, covers collision energy $\sqrt{s_{NN}}$ = 7.7--19.6\,GeV and is the estimated statistical uncertainty from the RHIC beam energy scan phase II program. As one can see, in order to complete the beam energy scan and determine if the QCD critical point exists, the CBM experiment at FAIR is necessary to fill the energy gap between \mbox{$\sqrt{s_{NN}}$ = 3--8 GeV~\cite{Almaalol:2022xwv}}; see the hatched region in Figure~\ref{BESI_netp}. Finally, it is worth noting that, in low-energy collisions, net-proton cumulants have also been recently proposed as a means for extracting the speed of sound and its logarithmic derivative \cite{Sorensen:2021zme}.

\subsection{Limits of Thermalization in High-Energy Nuclear Collisions}

In the previous subsection, the energy dependence of the high order moments of net-protons for the search of the QCD critical point in the beam energy scan program at RHIC is discussed. These high order moments, on the other hand, can also be used to test the limits of thermalization in such high-energy collisions~\cite{Gupta:2022phu}. In the past, yields of hadrons from high-energy nuclear collisions have been used to fit to the results of thermal model calculations in order to extract the freeze-out parameters; see, for example, discussions in Ref.~\cite{Andronic:2017pug}. The mean value of hadron yield is the first order of its multiplicity distributions. Certainly, the thermal analysis with the first moment $M_i$ is necessary, but not sufficient, for understanding the dynamics that lead to the macroscopic thermal behavior in high-energy nuclear collisions. Experimentals of high order moments have to be used in the analysis. Below, we report recent progress~\cite{Gupta:2011wh}.
 
In Equations~\eqref{eq:ratio0}--\eqref{eq:ratio2}, the terms on the left and in the middle are experimentally measured quantities while the generalized thermodynamic susceptibilities, shown on the right side of these equations, can be extracted from thermodynamic calculations with a given ensemble. By comparing the associated cumulant ratios with experimental data, one then can determine freeze-out temperature and chemical potential. In addition, the test of the limit of thermalization can be performed at any order where data are available. This approach has been used in the tests of Lattice QCD calculations~\cite{Gavai:2010zn,Bazavov:2012vg,Borsanyi:2014ewa,Bzdak:2012an,Cheng:2008zh,Bellwied:2013cta} and in a comparison with experimental results~\cite{Gupta:2011wh}.

\begin{figure}[H]
\includegraphics[width=0.94\textwidth]{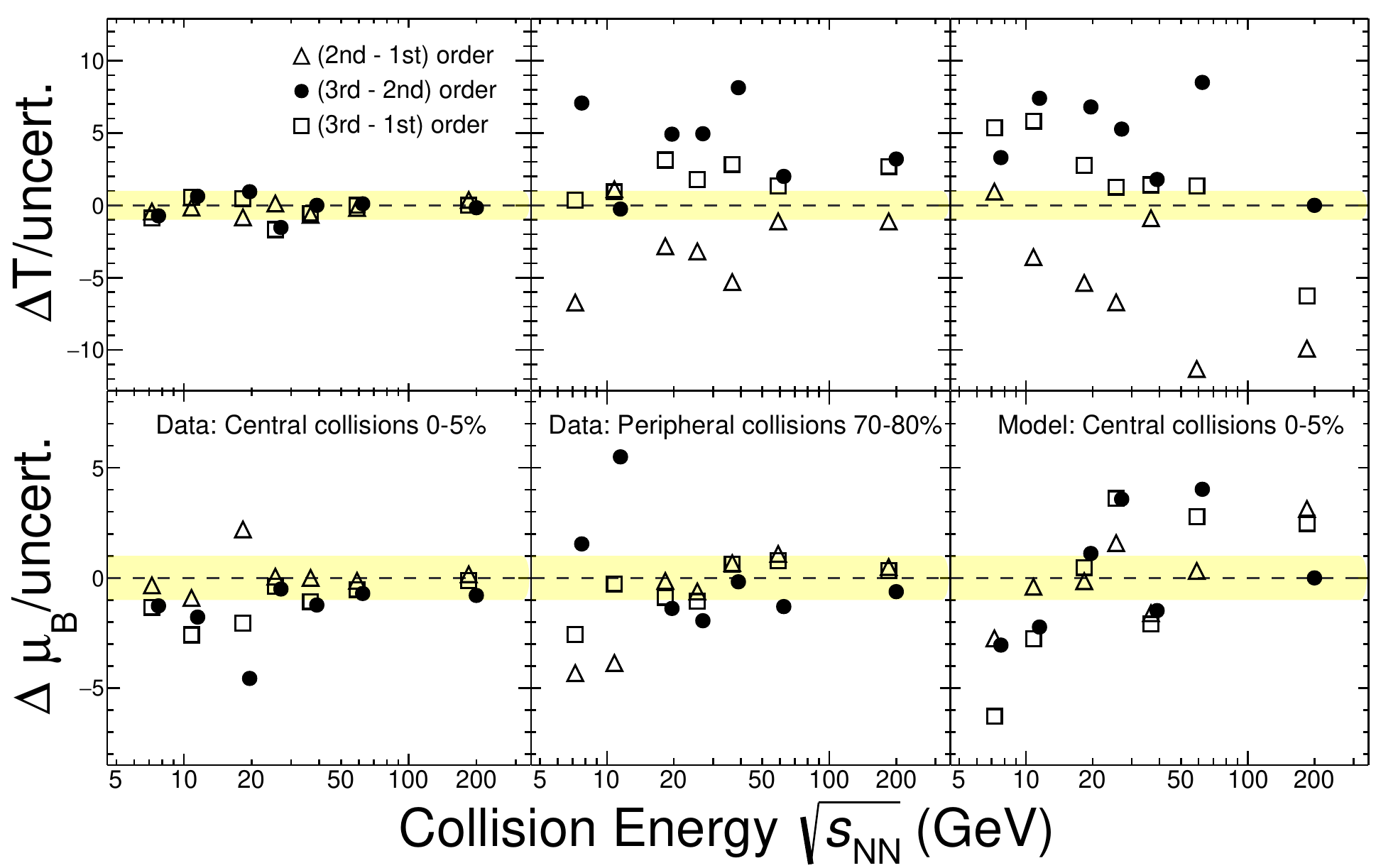}
\caption{Comparison of the freeze-out parameters obtained by fitting to different orders of high moments of net-proton, net-Kaon, and net-charge from RHIC  data~\cite{STAR:2010mib,STAR:2017tfy,STAR:2014egu}. The differences in freeze-out temperature ($\Delta T$) and baryon chemical potential ($\Delta \mu_B$) are shown in the top and bottom row, respectively. Triangles, dots, and squares represent the difference of net-proton cumulants between 2nd and 1st, 3rd and 2nd, and 3rd and 1st. Left and middle-columns are results from experimental data. The results from hadronic transport model UrQMD~\cite{Bass:1998ca,Bleicher:1999xi} simulations are shown in the right column. Yellow bands indicate 1$\sigma$ mark in differences.}
\label{fig:thrmal2} 
\end{figure}

Here, we present a study using STAR recent published results of net-proton, net-Kaon and net-charge~\cite{STAR:2010mib,STAR:2017tfy,STAR:2014egu} and compared to a thermal calculation~\cite{Gupta:2022phu} with grand canonical ensemble (GCE). The resulting differences in freeze-out temperature and chemical potential are displayed in Figure~\ref{fig:thrmal2}. Left- and middle-panels show the experimental data of central (0--5\%) and peripheral (70--80\%) Au+Au collisions. Right-panels display the results of the thermal comparison with central collisions generated by the hadronic transport model UrQMD~\cite{Bass:1998ca,Bleicher:1999xi}.  As one can see, neither peripheral data nor UrQMD central collisions is consistent with the GCE thermal model predictions for Au+Au collisions in the energy range from $\sqrt{s_{NN}}$ = 7.7--200 GeV. For central collisions, on the other hand, the differences in all orders are consistent within 1$\sigma$ in the high energy range $\sqrt{s_{NN}} \ge $ 40 GeV. The results imply that, up to the 3rd order, thermalization is indeed consistent with data in high-energy central heavy ion collisions, but the same cannot be said for collisions below the center of mass energy 40 GeV.  

\begin{figure}[H]
\includegraphics[width=0.83\textwidth]{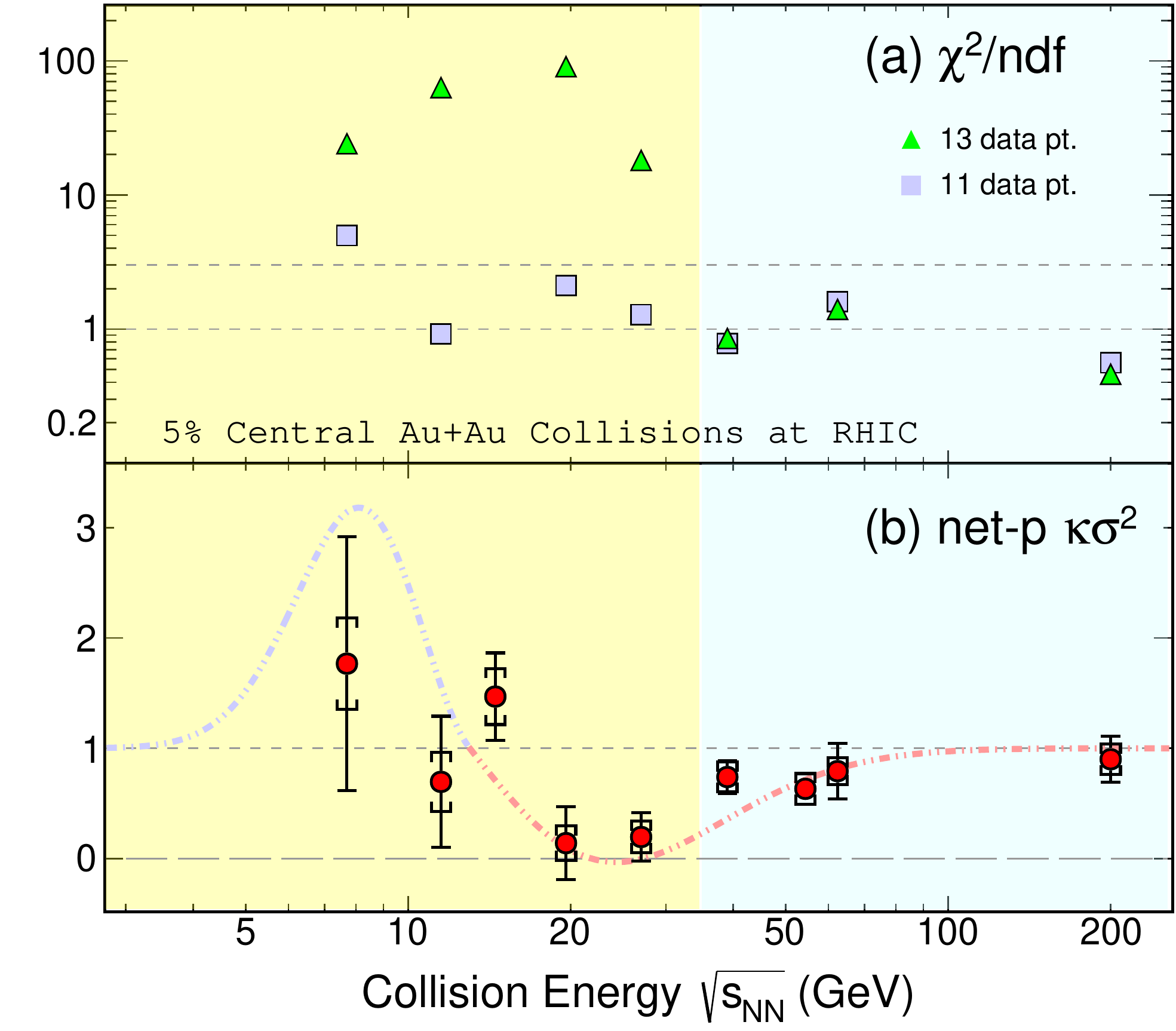}
 \caption{The top panel shows the energy dependence of $\chi^2/{\rm ndf}$ from the thermal fits with either 13~data points (green triangles) or 11 data points (gray squares). The bottom panel shows the energy dependence of the net-proton fourth-order cumulant ratios $C_4/C_2=\kappa\sigma^2$~\cite{STAR:2010mib}. The colour filling in both top and bottom panels tentatively divides the range of collision energies into regions that are clearly in agreement with the predictions of an ideal resonance gas, and therefore cannot contain the QCD critical point. Dot-dashed line is a fit result and indicates possible oscillation in the fourth-order net-proton cumulant ratios~\cite{Stephanov:2011pb}.  }
\label{fig:thrmal3} 
\end{figure}

The energy dependence of the net-proton fourth-order moment, $\kappa \sigma^2$ from central Au+Au collisions (low panel) are compared with the fitting results (top panel) in Figure~\ref{fig:thrmal3}. Similarly, at high energy, both fitting results shown in the top panel and net-proton kurtosis are consistent with the Grand Canonical limits, and clear deviations appear at energy lower than 40 GeV. In the net-proton kurtosis~\cite{STAR:2020tga}, the expected criticality pattern~\cite{Stephanov:2011pb}, as a function of energy, seems evident; see the dot-dashed line in the lower panel of Figure~\ref{fig:thrmal3}. It is necessary to point out that criticality is a non-equilibrium phenomenon. The observed deviation from Grand Canonical limit in low energy collisions could be caused by the system passing through the critical point, but it could also be that the system is out of equilibrium. The observation of departures from thermodynamic equilibrium in the final state opens up new directions in the study of heavy-ion collisions. Further analyzing the new experimental precision data~\cite{STAR:BESII2014} with Canonical Ensemble, where the effect of baryon number conservation is included, is necessary in order to understand the underlying physics in the finite baryon density region.

\section{Strange Quark Probes of Parton Dynamics and QCD Interactions}

Strange quark is a unique probe of the Quark--Gluon Plasma produced in high energy nucleus--nucleus collisions. Rafelski and Muller pointed out~\cite{Rafelski:1982pu,KOCH1986167} that, in a hot QGP with a temperature above 160 MeV, the strange quark with a mass of 100 MeV/c$^2$ can be abundantly produced via gluon--gluon fusion to a strange quark pair. Strangeness enhancement, an increased production of strange hadrons, multi-strange hyperons in particular, has been considered a signature of equilibrated Quark--Gluon Plasma (QGP). On the other hand, if a hadronic gas system can interact and maintain thermal equilibrium hypothetically for a sufficiently long time, strangeness production can also be significantly enhanced to reach an equilibrium. Therefore, it is important to investigate the underlying dynamics for strangeness production.

Strange hadrons can be identified by their decay topologies and such particle identification capability can be extended to momentum much higher than the limit from traditional time-of-flight detectors. Measurements of elliptic flow and nuclear modification factor of identified particles up to the intermediate transverse momentum region of 5 GeV/c have been essential in the discovery of the quark number scaling~\cite{star-parttype, star-multi-strange}. The quark number scaling is a manifestation of the quark coalescence dynamics for particle formation at the chemical freeze-out stage. The effective degrees of freedom before the hadronization must be dominated by parton dynamics, and the partonic hydrodynamics are responsible for the development of the azimuthal angular anisotropy in the flow measurements. The change of paradigm from a QGP of free quarks and gluons in a QCD bag to a QGP of almost perfect fluid of strongly interacting quarks and gluons is one of the most important achievements of the heavy ion collision physics in the past two decades. The coalescence dynamics showed that the partonic degrees of freedom dominate the evolution dynamics of the QCD matter created in high energy nucleus-nucleus collisions.

The $\phi$ meson and $\Omega$ hyperon have played a unique role in probing the partonic dynamics of the QGP. These strange particles do not have significant hadronic interaction cross sections with other hadrons during the hadronic evolution after the chemical freeze-out. Therefore, they can carry imprints from dynamics of the partonic phase of the QGP evolution. Using coalescence dynamics, these particles have been used to probe strange quark properties in the QGP at the chemical freeze-out \cite{phi-omega, star-phi-omega}. Such measurements have also been conducted with the STAR BES data to probe variations of the strange quark properties as the collision energy decreases~\cite{star-bes-strange}. We note that the ratio of $\Omega$ to $\bar{\Omega}$ increases as the colliding energy decreases from the STAR BES measurement, and the ratio is significantly above one at low energy. There is a net baryon number in $\Omega$ hyperons although strangeness conservation dictates that strange and anti-strange quarks must be produced in pairs. The dynamics of baryon number transport to $\Omega$ hyperons are a subject of interest. Measurements of correlations between $\Omega$ and other particles such as Kaon and anti-hyperons could shed light on the roles of strangeness and baryon number conservation in baryon transport dynamics~\cite{XTWu_sqm22}.   

The large number of strange hyperons produced in one single heavy ion collision also provided a unique opportunity to investigate hyperon--nucleon, hyperon--hyperon interactions through correlation measurements. The STAR experiment carried out the first meaningful measurement of a $\Lambda$--$\Lambda$ correlation function and extracted the interaction parameters between $\Lambda$s~\cite{star-L-L}. Motivated by calculations of the HAL QCD Collaboration~\cite{HALQCD-no}, STAR measured the $p$-$\Omega$ correlation function from Au+Au collisions~\cite{star-p-omega}. The STAR result, based on the ratio of correlations from two centrality bins as suggested by the theoretical calculation, slightly favors the existence of a bound $p$-$\Omega$ system. ALICE also measured correlations of $p$-$\Xi$ and $p$-$\Omega$ from $p+p$ collisions at 13 TeV energy. The ALICE data indicated a strong $p$-$\Omega$ attractive interaction though no signal was observed for the existence of a bound state~\cite{ALICE:2020mfd}. Existing measurements of hyperon correlations are not precise enough to provide definitive conclusions on the interaction strength between hyperons and the binding energy of possible composite hyperon states. The order of magnitude increase in the collision data samples in the coming years will enable us to utilize the heavy ion collisions as hyperon factories to experimentally address an important topic of hyperon interactions currently incomplete in QCD descriptions of strong interactions.

\begin{figure}[H]
\includegraphics[width=0.75\textwidth]{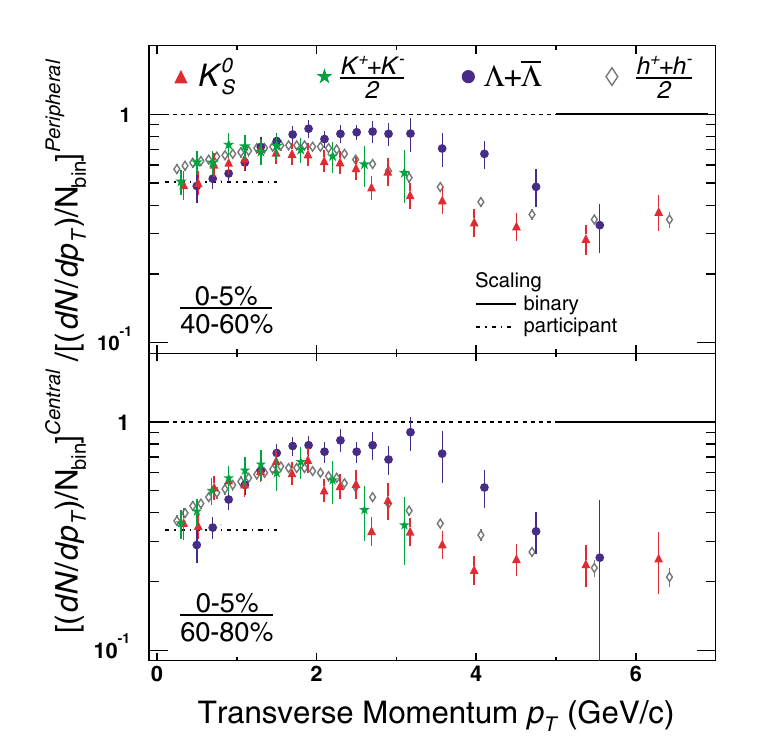}
\caption{Transverse momentum dependence of the nuclear modification factors for charged hadrons (open diamonds), $K_S^0$ (triangles), charged Kaons (stars) and $\Lambda$ hyperons (dots) from $\sqrt{s_{NN}} = 200$ GeV Au+Au collisions. Solid and dot-dashed lines stand for the expected scaling of binary collisions and number of participants, respectively. There is a clear particle-type (baryon versus meson) dependence below intermediate $p_T$ presumably due to hydrodynamic flow and coalescence dynamics. The particle-type dependence seems to disappear above $p_T$ of 6 GeV/c where partonic jet quenching dynamics may start to dominate.
 }
\label{Rcp-hyperon} 
\end{figure}

Because we can identify hyperons at high transverse momentum ($p_T$) region using their decay topologies, the particle-type (baryon versus meson) dependence of the nuclear modification factors can yield important insight on the parton energy loss in the QGP as well. Figure~\ref{Rcp-hyperon} shows the nuclear modification factors, $R_{\rm CP}$ ratio of normalized yields of central to peripheral collisions, as a function of  $p_T$ for charged hadrons, Kaons and $\Lambda$ hyperons~\cite{star-parttype}. The $R_{\rm CP}$ values for mesons and hyperons seem to approach each other around $p_T$ of 6~GeV/$c$. The disappearance of the particle-type dependence could be an important feature of the $R_{\rm CP}$: it would indicate a minimum $p_T$ above which the jet quenching dynamics may dominate, a landmark $p_T$ cut separating jet probe from hadrodynamical and coalescence regions. Improved measurements of the nuclear modification factors above this important $p_T$ cut of 6~GeV/c for identified particles in the next phase of RHIC scientific program from 2023--2025 could bring new insight on parton energy loss dynamics at RHIC.

\begin{figure}[H]
\includegraphics[width=0.87\columnwidth]{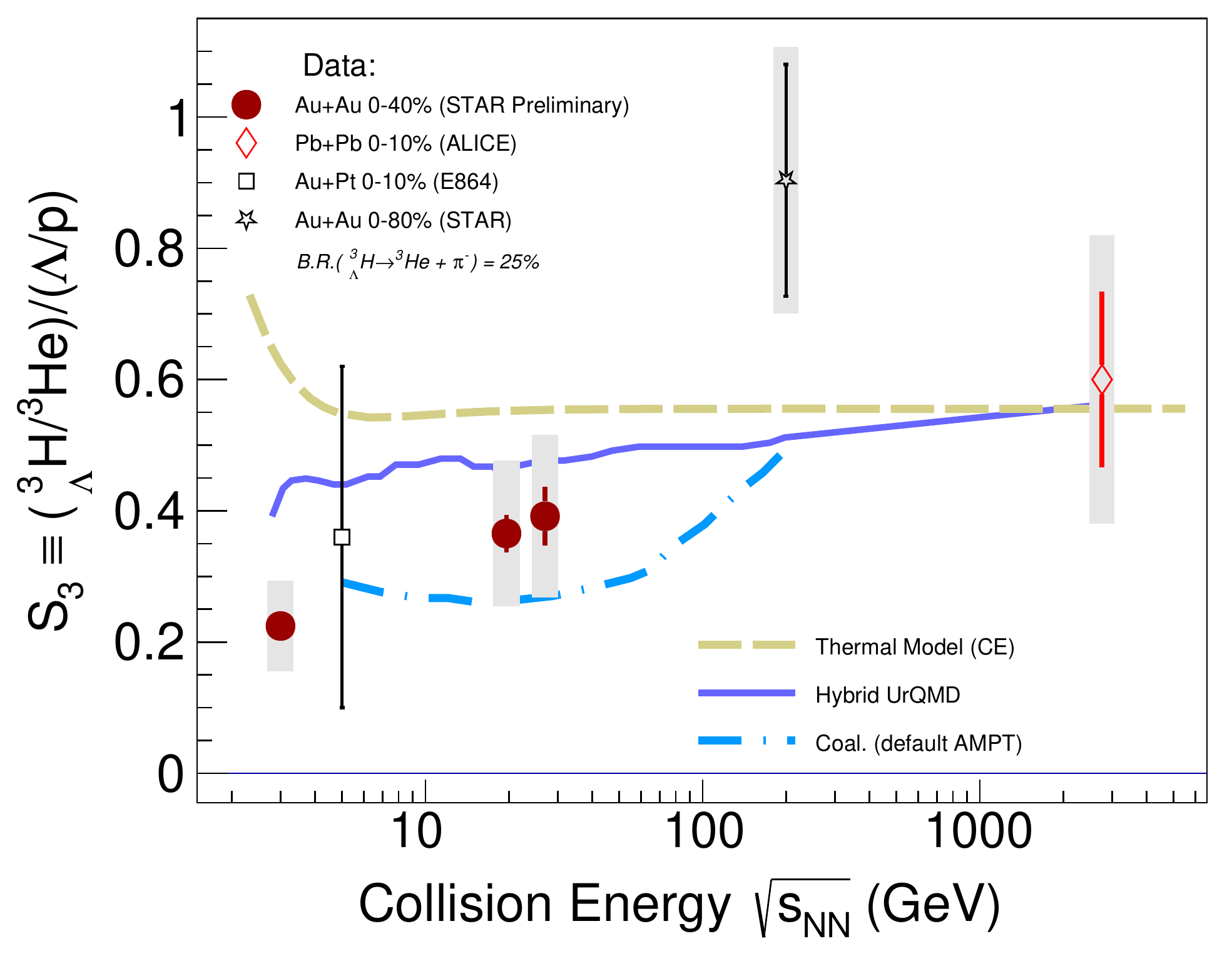} 
\caption{Strangeness population ratio $S_3$, from central heavy ion collisions, shown as a function of collision energy.  Model calculations by thermal model (gold dashed line), hybrid UrQMD (blue line), and AMPT with coalescence (dot-dashed line) are also presented. }\label{fig:S3strange}
\end{figure}

\textls[-15]{Nucleus--nucleus collisions also opened an important venue to study hyper-nuclei dynamics, especially from the STAR BES data to a lower energy regime at future facilities.
Driven by baryon density and the strangeness production threshold, both the thermal model~\cite{Andronic:2010qu,Andronic:2005yp}} and transport model~\cite{Steinheimer:2012tb} have predicted the production of light nuclei and hyper-nuclei to peak around $3 \le \sqrt{s_{NN}} \le 10$ GeV in high-energy nuclear collisions. To remove trivial factors including chemical potential and canonical effects, double ratios are often used;  
for instance, the ratio involving Hypertriton, Helium-3, $\Lambda$, and proton yields:
\begin{equation}
S_3=\frac{{^3_{\Lambda}H/^3He}}{\Lambda/p}\,.
\end{equation}
The experimental results of $S_3$ are shown in Figure~\ref{fig:S3strange} along with those from model calculations~\cite{Steinheimer:2012tb,Zhang:2009ba,Guo:2021udq,Ivanov:2020udj}. New precision data from STAR experiment show a gradual increase as a function of the collision energy, and the value seems to approach the equilibrium limit in collisions at the LHC~\cite{ALICE:2015oer,ALICE:2019vlx}. It is also interesting to note that the limiting value is about 2/3, which is commonly used in calculations of $\Lambda$--$N$ interactions~\cite{Glassel:2021rod}. In the high baryon density region where $\sqrt{s_{NN}}\le 10$ GeV, the double ratio is further away from the thermal limit, suggesting a clear density effect. More precision measurements are needed in order to understand the $Y$--$N$ and $Y$--$Y$ interactions as well as their implication to the inner structure of compact stars. For recent discussions on $Y$--$N$ and $Y$--$N$--$N$ interactions, see Ref.~\cite{Friedman:2022bor} and references therein. The next generation experiment CBM at FAIR with an unprecedentedly interaction rate capability will be very important for such measurements after RHIC. 

\section{Outlook: Physics at High Baryon Density}

\begin{figure}[H]
\includegraphics[width=0.82\textwidth]{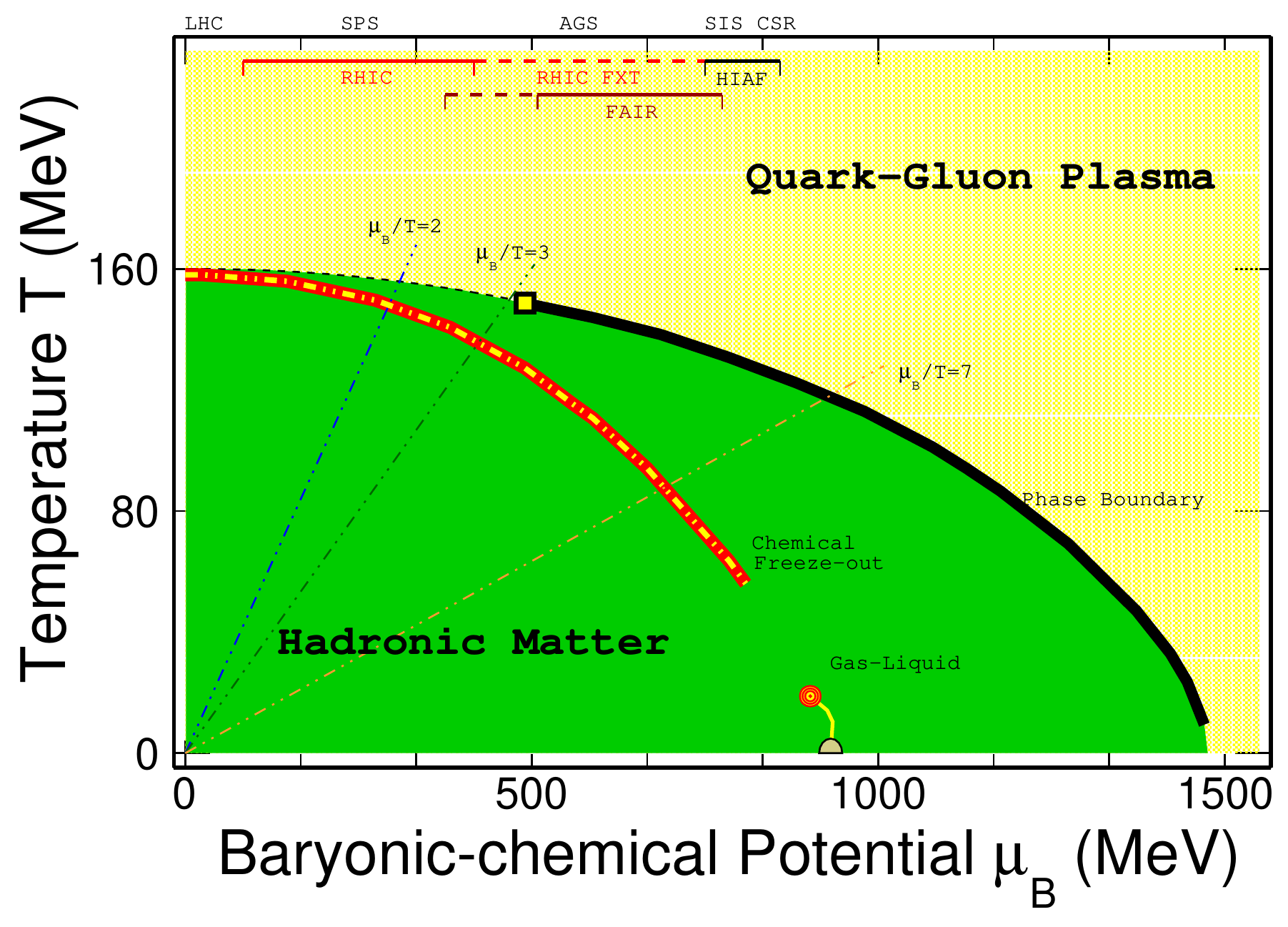}
\caption{Sketch of the QCD phase diagram. The dashed line represents the smooth crossover region up to $\mu_B/T \leq 3$. The black solid line represents the speculated first-order phase boundary. The empirical thermal freeze-out results from global hadron yield data are shown as the red-yellow line~\cite{Cleymans:2005xv}.  
 The liquid-gas transition region that features a second order critical point is shown by the red-circle, and a first-order transition line is shown by the yellow line, which connects the critical point to the ground state of nuclear matter. The coverage of the RHIC BES-II program ($\mu_B/T \sim 3$), STAR fixed target program ($\mu_B/T \sim 7$) and future FAIR and HIAF facilities are indicated at the top of the figure.}\label{fig1:phasestructure} 
\end{figure}

Since the discovery of the new form of matter, the strongly coupled Quark--Gluon Plasma (QGP)~\cite{STAR:2005gfr}, created in high-energy nuclear collisions in the early 2000s, scientists 
have been asking: ``What is the structure of the QCD phase diagram in the high baryon density region?'' and ``Is there a QCD critical point?''. Model studies have shown that a first-order phase boundary is expected at the finite baryon chemical potential $\mu_B$, while at vanishing $\mu_B$, there is a smooth crossover between QGP and hadronic matter. Thermodynamically, a critical point ought to be there at the end of the first-order phase boundary; see Figure~\ref{fig1:phasestructure}. More discussions on experimental results and Lattice calculations can be found in Refs.~\cite{Luo:2017faz,Bzdak:2019pkr,STAR:2021rls,Luo:2022mtp}.

The results, including the observations on collectivity, chirality, critical point, and strangeness production, from the first campaigning of the beam energy scan at RHIC show that partonic activities in central Au+Au collisions persist from $\sqrt{s_{NN}}=200$\, GeV to 39~GeV (corresponding to $\mu_B/T\leq3$) while in collisions at 3 GeV (corresponding to $\mu_B/T\sim7$) hadronic interactions dominate~\cite{star-multi-strange,phi-omega,STAR:2020tga,STAR:2021iop,STAR:2021yiu,STAR:2021fge}. In addition, charge separation, a measure of the strength of the effect of CME, is vanishing in low energy collisions~\cite{Adamczyk:2014mzf}. The QCD critical point, if existing, should be accessible in collisions at energy between \mbox{3--39 GeV}. RHIC's second phase energy scan has completed data taking, and more than 10~fold statistics of Au + Au collisions have been collected for energies between \mbox{7.7--19.6 GeV}. In order to complete the physics program of beam energy scan and study the phase structure of nuclear matter in high baryon density regions, future experiments such as CBM at FAIR are necessary. It will not only aid the search for the QCD critical point, but also extend the research to hyper-nuclear production which is important for studying the basic $Y-N$  interactions. The hyperon--nuclear interaction is one of the vital connections between the physics of high-energy nuclear collisions and inner structure of compact stars.  

\authorcontributions{All authors contributed equally. All authors have read and agreed to the published version of the manuscript.}

\funding{This work is supported in part by the National Key Research and Development Program of China under contract Nos. 2022YFA1604900, 2020YFE0202002, and 2018YFE0205201; 
the National Natural Science Foundation of China (NSFC) under contract Nos. 12122505, 12175084, 11890710(11890711), and 11835002; 
the U.S. Department of Energy (No. DE-SC0012910); and 
the Fundamental Research Funds for the Central Universities (CCNU220N003).}

\dataavailability{Not applicable.} 

\acknowledgments{We thank S. Gupta, V. Koch, D. Mallick, B. Mohanty, and M. Stephanov, for insightful discussions.}   

\conflictsofinterest{The authors declare no conflict of interest.} 



\begin{adjustwidth}{-\extralength}{0cm}

\reftitle{References}

\PublishersNote{}
\end{adjustwidth}
\end{document}